\documentclass[preprint,secnumarabic,amssymb, amsmath,nobibnotes, aps, prl, superscriptaddress,floatfix,showpacs]{revtex4-1}

\newcommand{\YBCO}{YBa$_2$Cu$_3$O$_{y}$ }

\usepackage{graphicx}
\usepackage[tight]{subfigure}
\usepackage{mathrsfs}
\begin{document}
\title{Magnetization of underdoped \YBCO above the irreversibility field}%

\author{Jing Fei Yu}%
\email[email: ]{jfeiyu@physics.utoronto.ca}
\affiliation{Department of Physics, University of Toronto, Toronto, Ontario, M5S 1A7, Canada}
\author{B. J. Ramshaw}
\affiliation{Los Alamos National Laboratory, Mail Stop E536, Los Alamos, New Mexico, 87545, USA}
\author{I. Kokanovi\'{c}}
\affiliation{Cavendish Laboratory, University of Cambridge, Cambridge, CB3 0HE, United Kingdom}
\affiliation{Department of Physics, Faculty of Science, University of Zagreb, P.O. Box 331, Zagreb, Croatia}
\author{K. A. Modic}
\author{N. Harrison}
\affiliation{Los Alamos National Laboratory, Mail Stop E536, Los Alamos, New Mexico, 87545, USA}
\author{James Day}
\affiliation{Department of Physics and Astronomy, University of British Columbia, Vancouver, British Columbia, V6T 1Z1, Canada}
\author{Ruixing Liang}
\author{W. N. Hardy}
\author{D. A. Bonn}
\affiliation{Department of Physics and Astronomy, University of British Columbia, Vancouver, British Columbia, V6T 1Z1, Canada}
\affiliation{Canadian Institute for Advanced Research, 180 Dundas St.W, Toronto, Ontario, M5S 1Z8, Canada}
\author{A. McCollam}
\affiliation{High Field Magnet Laboratory, Radboud University, Toernooiveld 7, 6525 ED Nijmegen, the Netherlands}
\author{S. R. Julian}
\affiliation{Department of Physics, University of Toronto, Toronto, Ontario, M5S 1A7, Canada}
\affiliation{Canadian Institute for Advanced Research, 180 Dundas St.W, Toronto, Ontario, M5S 1Z8, Canada}
\author{J. R. Cooper}
\affiliation{Cavendish Laboratory, University of Cambridge, Cambridge, CB3 0HE, United Kingdom}

\date{\today}
\begin{abstract} 
    Torque magnetization measurements on \YBCO (YBCO) at doping $y=6.67$($p=0.12$), in DC fields (\textit{B}) up to 33 T and temperatures down to 4.5 K, show that weak diamagnetism persists above the extrapolated irreversibility field $H_{\rm irr} (T=0) \approx 24$ T. The differential susceptibility $dM/dB$, however, is more rapidly suppressed for $B\gtrsim 16$ T than expected from the properties of the low field superconducting state, and saturates at a low value for fields $B \gtrsim 24$ T. In
    addition, torque measurements on a $p=0.11$ YBCO crystal in pulsed field up to 65 T and temperatures down to 8 K show similar behaviour, with no additional features at higher fields. We discuss several candidate scenarios to explain these observations: (a) superconductivity survives but is heavily suppressed at high field by competition with CDW order; (b) static superconductivity disappears near 24 T and is followed by a region of fluctuating superconductivity, which causes
    $dM/dB$ to saturate at high field; (c) the stronger 3D ordered CDW that sets in above 15 T may suppress the normal state spin susceptibility sufficiently to give an apparent diamagnetism of the magnitude observed.        
\end{abstract}
\maketitle
The possible existence of bulk superconductivity as $T\rightarrow0$ K above the irreversibility field ($H_{\rm irr}$)\footnote{For ease of comparison with Refs. \cite{Grissonnanche2014a} and \cite{Yu2014}, we use the same units (Tesla) and notation (e.g. $H_{\rm irr}$ and $H_{c2}$) throughout this paper.} in the cuprates has been a long standing question. 
Not only is this problem important for our understanding of the cuprates, but also because there is still debate\cite{Banerjee2013,Norman1996} about whether Cooper pairs persist in the region of the field-temperature plane where quantum oscillations are seen\cite{Doiron-Leyraud2007}.  

Many experimental efforts have been made to address this issue\cite{Li2010,Yu2014,Grissonnanche2014a,Riggs2011}. Diamagnetism has consistently been reported using torque magnetometry at high fields in many families of cuprates and it is argued that this observation shows the persistence of Cooper pairs above $H_{\rm irr}$\cite{Li2010}. For YBa$_2$Cu$_3$O$_{y}$, resistivity measurements have established $H_{\rm irr}(T = 0)$ to be below 30 T for fields along the \textit{c}-axis for dopings
between $p=0.11$ (OII) and $p=0.12$ (OVIII)\cite{Ramshaw2012a}. Moreover, X-ray\cite{Chang2012,Gerber2015}, NMR\cite{Wu2013}, and sound velocity measurements\cite{LeBoeuf2012} have demonstrated the existence of static charge density wave (CDW) order that competes with superconductivity: Ref. \cite{Gerber2015} shows a distinct long range 3D order that continues to grow at 28 T for an OVIII crystal. The CDW is strongest and the suppression of $H_{c2}$ is
largest at $p=0.125$ for
YBCO\cite{Chang2012, Hucker2014}. 

Recent thermal conductivity measurements by Grissonnanche \textit{et al. }\cite{Grissonnanche2014a} show a sharp transition precisely at the extrapolated $H_{\rm irr}(T=0)\simeq22$ T for OII YBCO. They have interpreted this feature (henceforth referred to as $H_K$) as a signature of $H_{c2}$, arguing that the end of the rapid rise in thermal conductivity at 22 T reflects a corresponding increase in the mean-free-path as a result of the sudden disappearance of vortices at
$H_{c2}$. On a crystal with same doping, Marcenat \textit{et al. }\cite{Marcenat2015} show that the specific heat saturates at a field $H_{cp}$. $H_{cp}(T)$ lies above $H_{\rm irr}(T)$, but they extrapolate to the same value at $T=0$ K\cite{Marcenat2015}. In contrast, torque measurements by F. Yu \textit{et al. }\cite{Yu2014} on a crystal with the same doping suggested diamagnetism persisting to fields much higher than
$H_K$. The debate is thus still open. 


\begin{figure}
    \includegraphics[width=240pt]{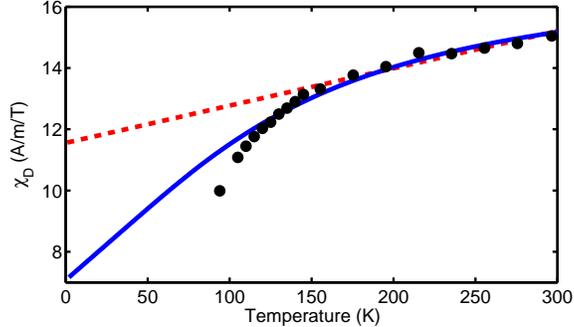}
    \caption{(Color online) \textit{Black dots}: high temperature anisotropic susceptibility $\chi_D(T)$ of the OVIII crystal at 10 T. 
        \textit{Blue solid line}: fit to this data above 120 K using Eq. \ref{eq:Eq1Kokanovic}. The parameters $A=11.09$ A/m/T, $T_F=680$ K are taken from Ref. \cite{Kokanovic2012}, while the fit gives $\chi^{VV} =5.84$ A/m/T , $T^* = 330$ K and $\chi^R(0) = 1.26$ A/m/T; \textit{Red dashed
            line}: Linear fit with $\chi(T) = 1.22\times10^{-2}\times(T+948)$, following Ref. \cite{Yu2014} but with different parameters. Note that $1\times 10^{-4}$ emu/mol$=9.73$ A/m/T.} 
    \label{fig:BG}
\end{figure}

To resolve this problem, we conducted torque magnetometry measurements of magnetization ($M$) on two $p=0.12$ (OVIII, $T_c = 65$ K) crystals in DC fields and one $p=0.11$ (OII, $T_c = 60$ K) crystal in pulsed fields. 
The crystals were mounted on piezoresistive cantilevers 
and placed on a rotating platform, with the CuO$_2$ planes parallel to the surface of the lever. DC field sweeps, first from 0 to 10 T and later from 0 to 33 T, were performed with the $c$-axis of the OVIII crystal at a small angle $\theta$ from the field. The magnetoresistance of the levers was eliminated by subtracting data from the complementary angle ($-\theta$). Similar procedures were used for the OII crystal in pulsed magnetic fields up to 65 T. For strongly anisotropic
superconductors, where out-of-plane screening currents can be neglected, the torque $\tau$ per unit volume $V$ at an angle $\theta$ from field $B$ is given by 
$\tau/V = \frac{1}{2}\chi_D(T)B^2\sin2\theta+M_cB\sin\theta$\cite{Kokanovic2013a}. Here $\chi_D(T) = \chi_c(T)-\chi_{ab}(T)$ is the anisotropic susceptibility in the normal state and $M_c$ is the magnetization from in-plane screening currents for a field of $B\cos\theta$ along the $c$-axis. This is a good approximation when $M_c \gg \chi_D B$ or when the superconducting gap and $M_c$ are both small.

A key challenge with magnetization measurements in the cuprates is the separation of the normal state from the superconducting contributions, because superconducting fluctuations are thought to contribute to $\chi(T)$ even at temperatures far above $T_c$\cite{Kokanovic2013a}, while $\chi^{\rm normal}$ is temperature dependent to well below $T_c$. We follow the procedure outlined in Refs.\cite{Kokanovic2013a,Kokanovic2012}
and interpret $\chi_D(T)$ in the normal state of underdoped YBCO as arising from the pseudogap and \textit{g}-factor anisotropy, plus a superconducting fluctuation term that sets in below 120 K. Neglecting isotropic Curie and core susceptibility terms, which do not contribute to $\tau$, the total normal state contribution to $\chi_D(T)$ is \cite{Kokanovic2012}: 
\begin{equation}
    \chi_D^{normal}(T) = \chi_D^{PG}(T)+\chi_D^{VV}+\chi_D^{R}(T)
    \label{eq:Eq1Kokanovic}
\end{equation}
where $\chi_D^{VV}$ is the $T$-independent Van Vleck susceptibility, $\chi_D^{PG}(T)$ is the pseudogap contribution assuming a V-shaped density of states (DOS)\cite{Loram1998}, 
and $\chi_D^{R}(T)$ is thought to arise from an electron pocket or Fermi arcs in the region $0.0184 < p <
0.135$. Specifically, $\chi_D^{PG} = A\left(1-y^{-1}\ln\left[\cosh(y)\right]\right)$,
where $A = N_0\mu_B^2$, $y = E_g/2k_BT$, $E_g$ = $k_BT^*$ and $T^*$ is the pseudogap temperature, and $\chi_D^{R}(T) = \chi^R(0)\left[1-\exp(-T_F/T)\right]$
where $T_F$ is the Fermi temperature.
The fit is shown in Fig. \ref{fig:BG}, along with a linear model for the normal state $\chi$ used in Ref. \cite{Yu2014}. Both fits agree well with the data for $T \geq 120$ K. Our background is almost twice as small as that of the linear fit at $T = 0$ K. Subtraction of the background magnetization using this non-linear model should thus give a significantly weaker diamagnetic signal at $T \rightarrow$ 0 K than the linear model would (about 160 A/m at 33 T).

\begin{figure}
    \subfigure[]{\label{fig:MagwithBG_a}\includegraphics[width=240pt]{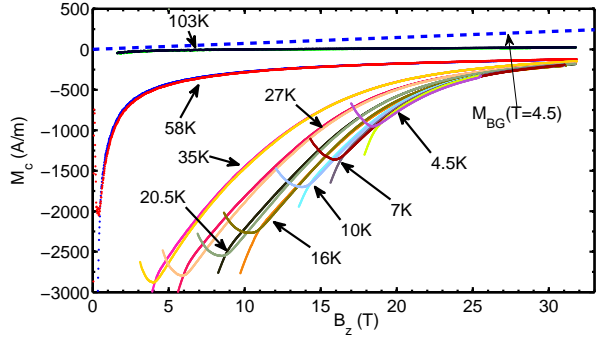}}
    \subfigure[]{\label{fig:MagwithBG_c}\includegraphics[width=240pt]{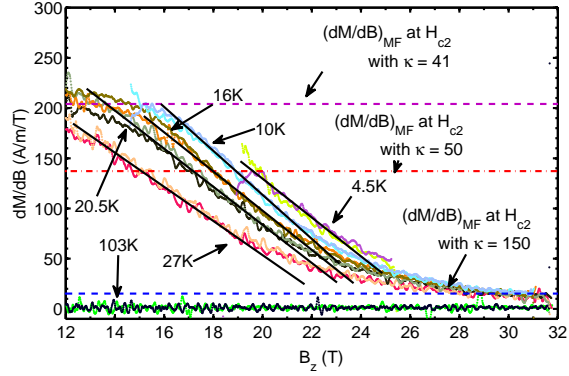}}%
    \caption{(Color online) \subref{fig:MagwithBG_a} Magnetization ($M_c$) of the OVIII crystal vs $B_z$, the field parallel to the \textit{c}-axis. Here $M_c(T,H) = M_{obs}(T,H)-M_{BG}(T,H)$, where $M_{BG} = \chi_D B$ and $\chi_D$ is the blue line in Fig. \ref{fig:BG}. \textit{Dashed line}: $M_{BG}$ at 4.5 K. Diamagnetism is present even at our highest field of 33 T. \subref{fig:MagwithBG_c} Differential susceptibility $dM/dB$ of the OVIII crystal vs $B_z$ at selected temperatures. The lines are guides to the eye. We call the characteristic field at which $dM/dB$
        departs from linearity $H_d$. \textit{Red}: calculated mean field $dM/dB$ near $H_{c2}$ with $\kappa=50$, with $\kappa=41$ (\textit{purple}) and with $\kappa=150$ (\textit{blue}).}
\end{figure}

\begin{figure}
   \subfigure[]{\label{fig:MagwithBG_b}\includegraphics[width=240pt]{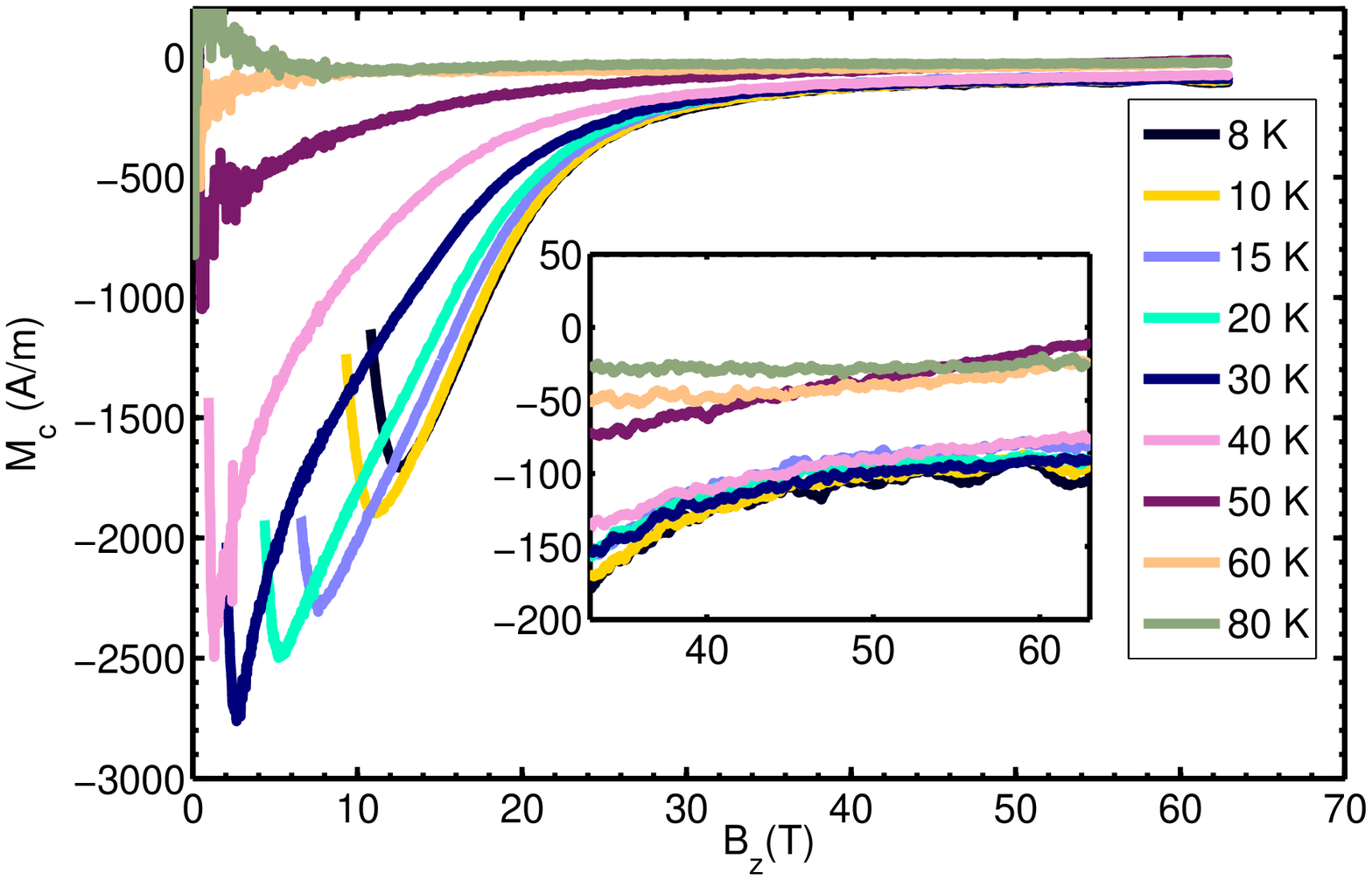}}
   \subfigure[]{\label{fig:MagwithBG_d}\includegraphics[width=240pt]{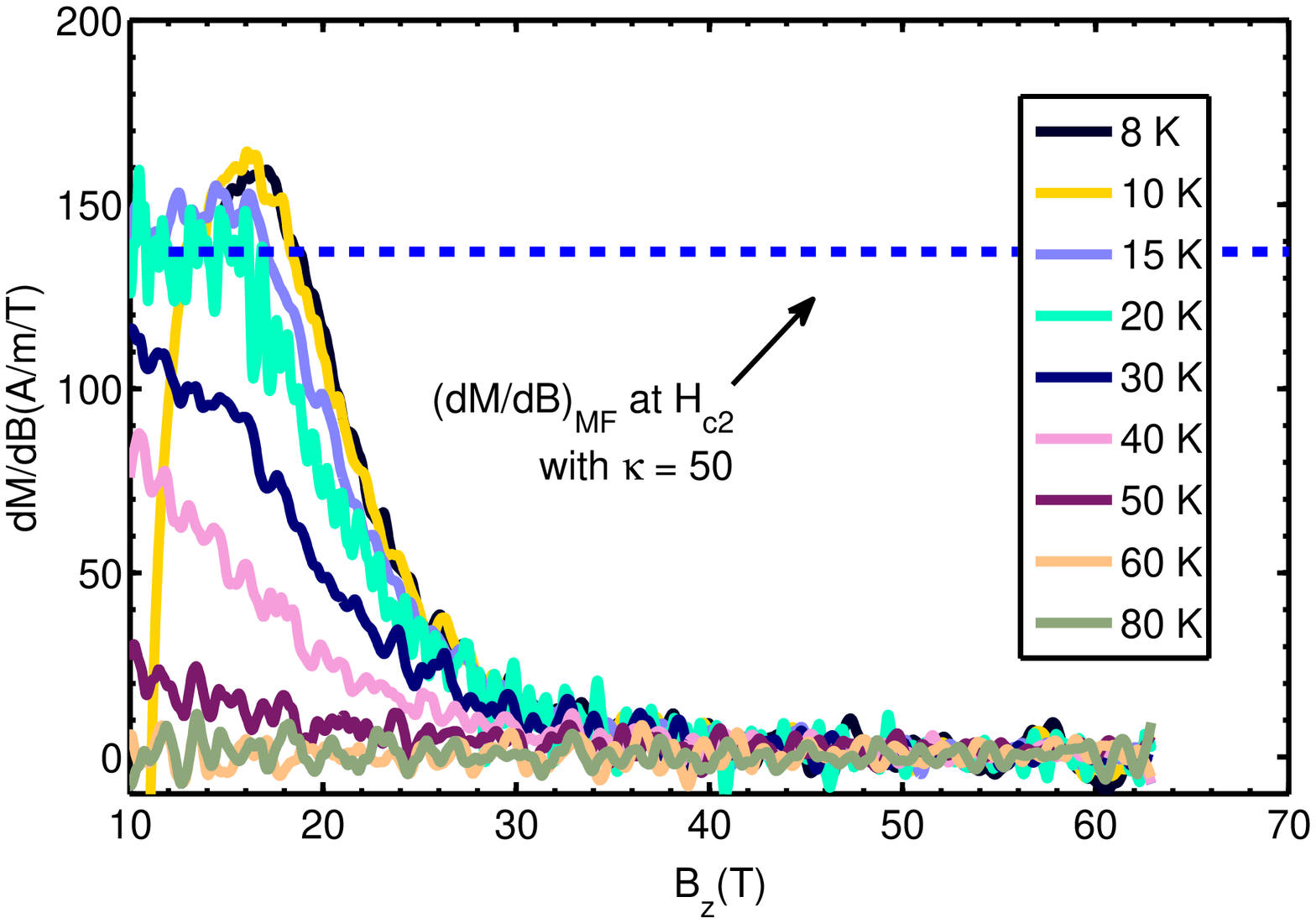}}
   \caption{(Color online) \subref{fig:MagwithBG_b} Magnetization ($M_c$) of the OII crystal measured in pulsed magnetic field up to $B_z=63$ T, where $M_c = M_{obs}-M_{BG}$, $M_{BG} = \chi_D B$ and $\chi_D$ is the blue line in Fig. \ref{fig:BG}. For clarity only the falling-field sweeps are shown. Diamagnetism is present though extremely weak at high field (inset). The small offset in $M_c$ between the $T \leq 40$ K and $T\geq 50$ K curves may be due to the transition to long-range CDW order near 40 K in high fields as observed in
       both sound velocity\cite{LeBoeuf2012} and NMR\cite{Wu2013}. \subref{fig:MagwithBG_d} Differential susceptibility for the OII crystal in pulsed field. $dM/dB$ is seen to be small and constant up to the highest field of 63 T. \textit{Blue}: calculated mean field $dM/dB$ near $H_{c2}$ with $\kappa=50$. } 
\end{figure}

In Figs. \ref{fig:MagwithBG_a} and \ref{fig:MagwithBG_b}, we show $M_c$ vs $B_z$ curves at selected temperatures for the OVIII and OII crystals, obtained by subtracting $M_{BG} = \chi_{BG}B$, where $\chi_{BG}$ is the blue line in Fig. \ref{fig:BG}, and $B_z$ is the field projected along the crystalline $c$-axis. For the OVIII crystal, at $T=103$ K, we see that $M_c$ is almost zero. At 58 K, just below $T_c$, we see significant diamagnetism that gradually tends to about
$-130$ A/m at high field. Fig. \ref{fig:MagwithBG_a} shows that the crystal remains weakly diamagnetic down to 4.5 K in fields up to 33 T. Similar behaviour was found for the OII crystal in pulsed fields. As shown in Fig. \ref{fig:MagwithBG_b}, $M_c$ is still diamagnetic at the highest field $B_z=63$ T, but has a small value -- about $-90$ A/m at 8 K. Our results differ from those of F. Yu \textit{et al. }\cite{Yu2014}: for example, at 10 K and 20 T we measure $M_c$ to be four times
larger\footnote{For OII crystal, the calibration factor used to obtain the absolute value of $M_c$ was found by requiring the $B^2$ term in the torque data ($\tau/V = 1/2\chi B^2\sin2\theta$) between 20 and 64 T at 80 K to give the value $\chi_D(80) = 11.2$ A/m/T on the blue line in Fig. \ref{fig:BG}. For OVIII, excellent $\sin2\theta$ behaviour was observed in the torque at 10 T and 97 K at the HFML, and was then normalized to the measured value of $\chi_D(97)$ obtained from
    angle sweeps on the same crystal in the Cavendish Laboratory at Cambridge, whose results are shown by solid points in Fig. \ref{fig:BG}. In the present high field experiments the crystals had to be small, so the $T$-dependence of the lever sensitivity could not be found from the gravitational torque in the usual way. It was obtained from the consistent normalized $T$-dependences of several levers of the same type in zero
field gravitational torque measurements on larger crystals.}; furthermore, at 30 T
we find about -200 A/m for OII and OVIII rather than their value of -75 A/m. Our estimated uncertainty in $\chi_D(0)$
corresponds to $\pm 32$ A/m in $M_c$ at 33 T and $\pm 61$ A/m at 63 T. 
%

Although the weak diamagnetic signal persists to higher fields, we are able to see a signature in our differential susceptibility $dM/dB$ at fields comparable to $H_K$ (22 T) found by thermal conductivity\cite{Grissonnanche2014a}. In each curve of Fig. \ref{fig:MagwithBG_c} and \ref{fig:MagwithBG_d}, $dM/dB$ decreases linearly, up to a field we call $H_d(T)$, before saturating to a small but non-zero value. 
At the lowest temperatures for both OVIII and OII crystals, we find $H_d\approx 24$ T, which is close to the extrapolated $H_{\rm irr}(T=0)$. This is consistent with the feature at $H_K$ found by thermal conductivity\cite{Grissonnanche2014a}, though unlike $H_K$, $H_d$ does not correspond to a sharp transition. $H_d$ varies very little with temperature for $T<10$ K, a result that is consistent with the findings of Ref.
\cite{Grissonnanche2014a}, though the $T$-dependence at high temperatures is not consistent with that found by Refs. \cite{Marcenat2015,Yu2014}. Surprisingly, we do not observe in any of our crystals the broad peak in $dM/dB$ reported by Ref. \cite{Yu2014}. 

\begin{figure}
    \includegraphics[width=240pt]{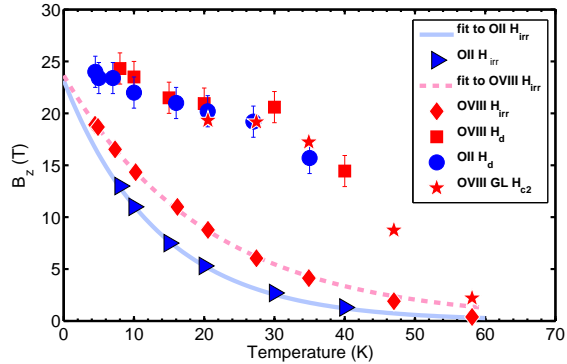}
    \caption{(Color online) $H_d$ for both OII and OVIII crystals show similar temperature dependences. Exponential fits to $H_{\rm irr}$ of OII($23.2\exp(-T/13.5)$)) and OVIII($23.7\exp(-T/20.5)$) give extrapolated values $H_{\rm irr}(0)$ = 23.2 and 23.7 T. These values are close to the low temperature $H_d$ for both crystals. Note that $H_{c2}$ from GL fits (see main text) connects smoothly to $H_d$. }
    \label{fig:Hdepart_and_Hirr}
\end{figure}

In highly anisotropic type-II superconductors, the magnetization calculated using mean field (MF) Ginzburg-Landau (GL) theory for an $s$-wave superconductor, which we use in the absence of a \textit{d}-wave theory, yields logarithmic behaviour at low field (in cgs units), $-4\pi M = \alpha \phi_0/(8\pi\lambda^2)\ln(\beta H_{c2}/H)$ for $0.02 < H/H_{c2} < 0.3$, where $\alpha$ and $\beta$ are numbers of order 1, $\phi_0$ is the flux quantum for Cooper pairs and $\lambda$ is the London
penetration depth\cite{Hao1991a}. $\mu$SR at low fields has shown a $\sqrt H$ field dependence for $\lambda (T = 0)$\cite{Sonier2007}, but results of tunnelling experiments on Bi-2212 imply thermally induced pair breaking near the nodes\cite{Benseman2015}, indicating a weaker field dependence at higher $T$. Thus, for simplicity, we assume a negligible field dependence of $\lambda$. We also assume $\alpha = 0.77$ and $\beta
= 1.44$  for $0.02 < H/H_{c2} < 0.3$\cite{Hao1991a}, in reasonable agreement with later works\cite{Pogosov2000,Bosma2011}, and we fit the low field
magnetization and obtain an estimate of $H_{c2}(T)$, shown in Fig. \ref{fig:Hdepart_and_Hirr}.  Since our GL values of $H_{c2}$ join smoothly to $H_d$, it is possible to interpret $H_d$ as the low temperature GL type $H_{c2}$. 

\begin{figure}
    \includegraphics[width=240pt]{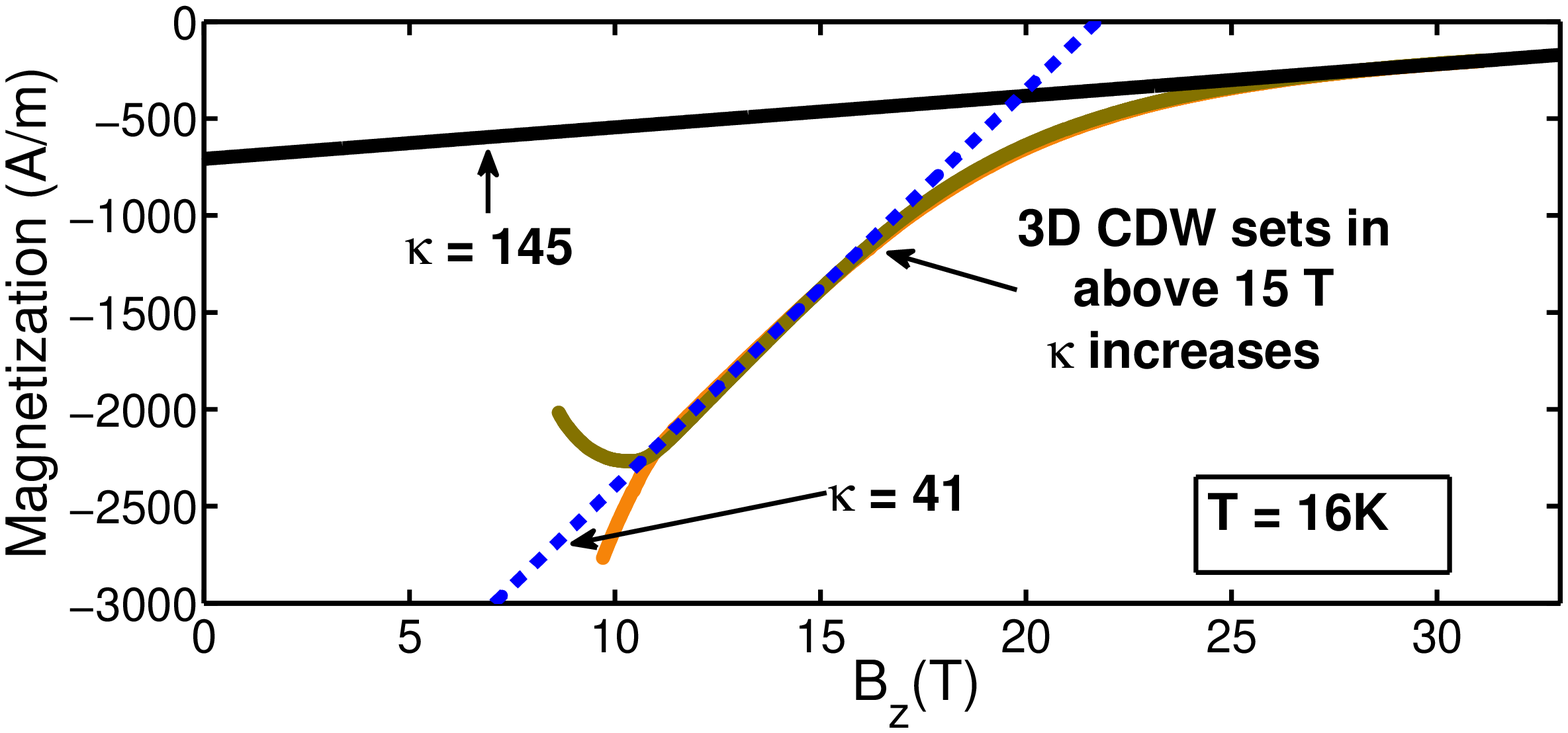}
    \caption{ Magnetization data of the OVIII crystal at 16 K. The blue dashed line shows the MF behaviour near $H_{c2}$ for an \textit{s}-wave superconductor with $\kappa=41$. The stronger (3D) CDW sets in above 15 T for OVIII YBCO. At higher fields the data are consistent with $\kappa=145$(\textit{solid line}).} 
    \label{fig:kappa_crossover_and_fluc}
\end{figure}

When $H/H_{c2} > 0.3$, and again using cgs units for an $s$-wave superconductor, the magnetization is expected to obey $4\pi M=(H-H_{c2})/[(2\kappa^2-1)\beta_A]$, where $\kappa$ is the GL parameter and $\beta_A = 1.16$ is the Abrikosov parameter\cite{Tinkham1996,Hao1991}. Figs. \ref{fig:MagwithBG_c} and \ref{fig:MagwithBG_d} show that for $B > 28$ T, $dM/dB$ has the mean field property of saturating toward a constant value, but this
is very small and requires $\kappa \simeq 150$, a value far greater than $\kappa=50$ given Ref. \cite{Grissonnanche2014a}. This means that our high field $dM/dB$ is nearly ten times smaller than would be expected. This may be due to the field dependent charge density wave (CDW) order within the vortex liquid region\cite{Chang2012,Gerber2015}. The CDW competes with superconductivity and is partially suppressed at low field. As increasing field
suppresses superconductivity, the CDW order is gradually restored\cite{LeBoeuf2012}. The presence of a relatively strong CDW would increase $\lambda$ and thus increase $\kappa$, as illustrated in Fig. \ref{fig:kappa_crossover_and_fluc}. A linear region in $M_c(B)$ can also be seen in Fig. \ref{fig:MagwithBG_a}, for $T = 20$ K and $T=16$ K and $B\leq17$ T, with $\kappa = 41$, and in Fig. \ref{fig:MagwithBG_b}, for $T = 20$ K and $B\leq17$ T, with $\kappa=50$. 
These linear regions are not present above 20 K, where $M_c(B)$ is likely to be smeared out by thermal fluctuations. As shown in Fig. \ref{fig:kappa_crossover_and_fluc}, for the OVIII crystal, low field $M_c$ extrapolates to zero around 24 T, consistent with our GL type $H_{c2}$. This is the first time that clear linear behaviour, with the expected values of $\kappa$, has been observed in hole-doped cuprates. 

The value of $H_{c2}(0) \approx H_d \approx 24$ T obtained from these GL analyses may refer to a low field, unreconstructed Fermi surface. For fields greater than 24 T, we may be observing MF behaviour of weak superconductivity arising from the small electron pockets\cite{Doiron-Leyraud2007,LeBoeuf2007} resulting from the appearance of CDW order. The GL type theory we applied assumes \textit{s}-wave superconductivity and we cannot rule out
possible \textit{d}-wave effects on the determination of $H_{c2}$. An obvious possibility is the Volovik effect whereby the Cooper pairs near the nodes on the Fermi surface are broken up, and consequently, $\lambda$ and $\kappa$ would increase.

Alternatively, the diamagnetism that we observe above 24 T could be caused by superconducting fluctuations. The OII data in the insert to Fig. \ref{fig:MagwithBG_b} show that it is $\sim -100$ A/m between 35 and 63 T. This is 5 times smaller and falls more quickly with field than predicted by theory\cite{Galitski2000} for a 2D \textit{s}-wave superconductor at low temperatures and high fields. This is a robust statement because in the clean
limit all parameters in the theoretical expression\cite{Galitski2000} for $M_c(B)$ above $H_{c2}$ are known. Nernst data\cite{Chang2011} for OVIII crystal show saturation near 30 T to the negative value expected for an electron pocket. This does not necessarily rule out bulk superconductivity above 30 T because in the presence for a CDW, the vortex core entropy -- which dominates the Nernst effect -- could be reduced. However at a qualitative level, the Nernst data between
24 and 30 T may be more consistent with superconducting fluctuations. Since torque magnetization is sensitive to superconducting fluctuations while thermal conductivity sees only the normal quasi-particles which are the only source of entropy, this may explain why we do not observe the sharp transition at $H_K$ seen in Ref. \cite{Grissonnanche2014a}.  

Finally, the diamagnetism of $-90$ A/m observed at 63 T might be caused by a reduction in spin susceptibility associated with the stronger (3D) CDW order that sets in above 15 T\cite{Gerber2015}. The change required would be $1.36$ A/m/T in $\chi_D(0)$. This is consistent with the significant decrease in diamagnetism between 40 and 50 K shown in the inset of Fig. \ref{fig:MagwithBG_b}, the region where the 3D CDW seen at high fields goes away\cite{Gerber2015}. 

The above discussion highlights the importance of competing CDW and superconductivity instabilities\cite{Hayward2014a,Chang2012}. Little is known about the size of the CDW energy gap, or the MF behaviour expected for a $d$-wave superconductor just below $H_{c2}$ as $T \rightarrow0$ K. Therefore the linear $H$ dependence of
$dM/dB$ we observe below $H_d$ might be a fundamental property of a $d$-wave superconductor. In other words, because of Volovik-type pair breaking effects, the MF transition at $H_{c2}$ could have a discontinuity in $d^2M/dB^2$, rather than in $dM/dB$, which is the standard MF result for the second order transition in a conventional \textit{s}-wave superconductor. 

In summary, we observe diamagnetism in OVIII YBCO at fields up to 33 T and OII YBCO at fields up to 65 T using torque magnetometry. The analysis uses a different model for the high temperature normal state susceptibility that gives a smaller correction at low temperature compared with earlier models. 
We also find that $dM/dB$ departs from a linear lower field behaviour at fields $H_d \approx H_{\rm irr}(0) \approx 24$ T, and approaches a constant value at higher fields. We propose several candidate scenarios: a competing order scenario where a fully-fledged CDW at high field mostly suppresses the superconductivity so that the diamagnetism at high field could be attributed to bulk superconductivity; or a fluctuation picture in
which for $H > H_d$, the system crosses over to superconducting fluctuation behaviour; and lastly, the apparent diamagnetism could reflect a reduction in spin susceptibility associated with the stronger 3D CDW order. It would be of interest to develop $d$-wave expressions for the MF magnetization and for the fluctuation contribution in the low temperature, high field regime, for comparison with our data. This could settle the debate over the existence of the high field vortex liquid region.  

We thank G. Grissonnanche for useful discussions. This work was generously supported by NSERC and CIFAR of Canada, Canada Research Chair, EPSRC (UK) under Grant No. EP/K016709/1, Croatian Science Foundation project (No. 6216) and the Croatian Research Council, MZOS NEWFELPRO project No. 19. We thank HFML-RU, a member of the European Magnetic Field Laboratory. The work at LANL was funded by the Department of Energy Basic Energy Sciences program `Science at 100 T'.  The NHMFL
facility is funded by the Department of Energy, the State of Florida, and the NSF under cooperative agreement DMR-1157490.

\bibliography{paperBib}

\begin{thebibliography}{30}%
\makeatletter
\providecommand \@ifxundefined [1]{%
 \@ifx{#1\undefined}
}%
\providecommand \@ifnum [1]{%
 \ifnum #1\expandafter \@firstoftwo
 \else \expandafter \@secondoftwo
 \fi
}%
\providecommand \@ifx [1]{%
 \ifx #1\expandafter \@firstoftwo
 \else \expandafter \@secondoftwo
 \fi
}%
\providecommand \natexlab [1]{#1}%
\providecommand \enquote  [1]{``#1''}%
\providecommand \bibnamefont  [1]{#1}%
\providecommand \bibfnamefont [1]{#1}%
\providecommand \citenamefont [1]{#1}%
\providecommand \href@noop [0]{\@secondoftwo}%
\providecommand \href [0]{\begingroup \@sanitize@url \@href}%
\providecommand \@href[1]{\@@startlink{#1}\@@href}%
\providecommand \@@href[1]{\endgroup#1\@@endlink}%
\providecommand \@sanitize@url [0]{\catcode `\\12\catcode `\$12\catcode
  `\&12\catcode `\#12\catcode `\^12\catcode `\_12\catcode `\%12\relax}%
\providecommand \@@startlink[1]{}%
\providecommand \@@endlink[0]{}%
\providecommand \url  [0]{\begingroup\@sanitize@url \@url }%
\providecommand \@url [1]{\endgroup\@href {#1}{\urlprefix }}%
\providecommand \urlprefix  [0]{URL }%
\providecommand \Eprint [0]{\href }%
\providecommand \doibase [0]{http://dx.doi.org/}%
\providecommand \selectlanguage [0]{\@gobble}%
\providecommand \bibinfo  [0]{\@secondoftwo}%
\providecommand \bibfield  [0]{\@secondoftwo}%
\providecommand \translation [1]{[#1]}%
\providecommand \BibitemOpen [0]{}%
\providecommand \bibitemStop [0]{}%
\providecommand \bibitemNoStop [0]{.\EOS\space}%
\providecommand \EOS [0]{\spacefactor3000\relax}%
\providecommand \BibitemShut  [1]{\csname bibitem#1\endcsname}%
\let\auto@bib@innerbib\@empty
\bibitem [{Note1()}]{Note1}%
  \BibitemOpen
  \bibinfo {note} {For ease of comparison with Refs. \cite {Grissonnanche2014a}
  and \cite {Yu2014}, we use the same units (Tesla) and notation (e.g.
  $H_{\protect \rm irr}$ and $H_{c2}$) throughout this paper.}\BibitemShut
  {Stop}%
\bibitem [{\citenamefont {Banerjee}\ \emph {et~al.}(2013)\citenamefont
  {Banerjee}, \citenamefont {Zhang},\ and\ \citenamefont
  {Randeria}}]{Banerjee2013}%
  \BibitemOpen
  \bibfield  {author} {\bibinfo {author} {\bibfnamefont {S.}~\bibnamefont
  {Banerjee}}, \bibinfo {author} {\bibfnamefont {S.}~\bibnamefont {Zhang}}, \
  and\ \bibinfo {author} {\bibfnamefont {M.}~\bibnamefont {Randeria}},\ }\href
  {\doibase 10.1038/ncomms2667} {\bibfield  {journal} {\bibinfo  {journal}
  {Nature Communications}\ }\textbf {\bibinfo {volume} {4}},\ \bibinfo {pages}
  {1700} (\bibinfo {year} {2013})}\BibitemShut {NoStop}%
\bibitem [{\citenamefont {Norman}\ and\ \citenamefont
  {MacDonald}(1996)}]{Norman1996}%
  \BibitemOpen
  \bibfield  {author} {\bibinfo {author} {\bibfnamefont {M.~R.}\ \bibnamefont
  {Norman}}\ and\ \bibinfo {author} {\bibfnamefont {A.~H.}\ \bibnamefont
  {MacDonald}},\ }\href {\doibase 10.1103/PhysRevB.54.4239} {\bibfield
  {journal} {\bibinfo  {journal} {Physical Review B}\ }\textbf {\bibinfo
  {volume} {54}},\ \bibinfo {pages} {4239} (\bibinfo {year}
  {1996})}\BibitemShut {NoStop}%
\bibitem [{\citenamefont {Doiron-Leyraud}\ \emph {et~al.}(2007)\citenamefont
  {Doiron-Leyraud}, \citenamefont {Proust}, \citenamefont {LeBoeuf},
  \citenamefont {Levallois}, \citenamefont {Bonnemaison}, \citenamefont
  {Liang}, \citenamefont {Bonn}, \citenamefont {Hardy},\ and\ \citenamefont
  {Taillefer}}]{Doiron-Leyraud2007}%
  \BibitemOpen
  \bibfield  {author} {\bibinfo {author} {\bibfnamefont {N.}~\bibnamefont
  {Doiron-Leyraud}}, \bibinfo {author} {\bibfnamefont {C.}~\bibnamefont
  {Proust}}, \bibinfo {author} {\bibfnamefont {D.}~\bibnamefont {LeBoeuf}},
  \bibinfo {author} {\bibfnamefont {J.}~\bibnamefont {Levallois}}, \bibinfo
  {author} {\bibfnamefont {J.-B.}\ \bibnamefont {Bonnemaison}}, \bibinfo
  {author} {\bibfnamefont {R.}~\bibnamefont {Liang}}, \bibinfo {author}
  {\bibfnamefont {D.~A.}\ \bibnamefont {Bonn}}, \bibinfo {author}
  {\bibfnamefont {W.~N.}\ \bibnamefont {Hardy}}, \ and\ \bibinfo {author}
  {\bibfnamefont {L.}~\bibnamefont {Taillefer}},\ }\href {\doibase
  10.1038/nature05872} {\bibfield  {journal} {\bibinfo  {journal} {Nature}\
  }\textbf {\bibinfo {volume} {447}},\ \bibinfo {pages} {565} (\bibinfo {year}
  {2007})}\BibitemShut {NoStop}%
\bibitem [{\citenamefont {Li}\ \emph {et~al.}(2010)\citenamefont {Li},
  \citenamefont {Wang}, \citenamefont {Komiya}, \citenamefont {Ono},
  \citenamefont {Ando}, \citenamefont {Gu},\ and\ \citenamefont
  {Ong}}]{Li2010}%
  \BibitemOpen
  \bibfield  {author} {\bibinfo {author} {\bibfnamefont {L.}~\bibnamefont
  {Li}}, \bibinfo {author} {\bibfnamefont {Y.}~\bibnamefont {Wang}}, \bibinfo
  {author} {\bibfnamefont {S.}~\bibnamefont {Komiya}}, \bibinfo {author}
  {\bibfnamefont {S.}~\bibnamefont {Ono}}, \bibinfo {author} {\bibfnamefont
  {Y.}~\bibnamefont {Ando}}, \bibinfo {author} {\bibfnamefont {G.~D.}\
  \bibnamefont {Gu}}, \ and\ \bibinfo {author} {\bibfnamefont {N.~P.}\
  \bibnamefont {Ong}},\ }\href {\doibase 10.1103/PhysRevB.81.054510} {\bibfield
   {journal} {\bibinfo  {journal} {Physical Review B}\ }\textbf {\bibinfo
  {volume} {81}},\ \bibinfo {pages} {054510} (\bibinfo {year}
  {2010})}\BibitemShut {NoStop}%
\bibitem [{\citenamefont {Yu}\ \emph {et~al.}()\citenamefont {Yu},
  \citenamefont {Hirschberger}, \citenamefont {Loew}, \citenamefont {Li},
  \citenamefont {Lawson}, \citenamefont {Asaba}, \citenamefont {Kemper},
  \citenamefont {Liang}, \citenamefont {Porras}, \citenamefont {Boebinger},
  \citenamefont {Singleton}, \citenamefont {Keimer}, \citenamefont {Li},\ and\
  \citenamefont {Ong}}]{Yu2014}%
  \BibitemOpen
  \bibfield  {author} {\bibinfo {author} {\bibfnamefont {F.}~\bibnamefont
  {Yu}}, \bibinfo {author} {\bibfnamefont {M.}~\bibnamefont {Hirschberger}},
  \bibinfo {author} {\bibfnamefont {T.}~\bibnamefont {Loew}}, \bibinfo {author}
  {\bibfnamefont {G.}~\bibnamefont {Li}}, \bibinfo {author} {\bibfnamefont
  {B.~J.}\ \bibnamefont {Lawson}}, \bibinfo {author} {\bibfnamefont
  {T.}~\bibnamefont {Asaba}}, \bibinfo {author} {\bibfnamefont {J.~B.}\
  \bibnamefont {Kemper}}, \bibinfo {author} {\bibfnamefont {T.}~\bibnamefont
  {Liang}}, \bibinfo {author} {\bibfnamefont {J.}~\bibnamefont {Porras}},
  \bibinfo {author} {\bibfnamefont {G.~S.}\ \bibnamefont {Boebinger}}, \bibinfo
  {author} {\bibfnamefont {J.}~\bibnamefont {Singleton}}, \bibinfo {author}
  {\bibfnamefont {B.}~\bibnamefont {Keimer}}, \bibinfo {author} {\bibfnamefont
  {L.}~\bibnamefont {Li}}, \ and\ \bibinfo {author} {\bibfnamefont {N.~P.}\
  \bibnamefont {Ong}},\ }\href {http://arxiv.org/abs/1402.7371} {\ }\Eprint
  {http://arxiv.org/abs/1402.7371} {arXiv:1402.7371} \BibitemShut {NoStop}%
\bibitem [{\citenamefont {Grissonnanche}\ \emph {et~al.}(2014)\citenamefont
  {Grissonnanche}, \citenamefont {Cyr-Choini\`{e}re}, \citenamefont
  {Lalibert\'{e}}, \citenamefont {{Ren\'{e} de Cotret}}, \citenamefont
  {Juneau-Fecteau}, \citenamefont {Dufour-Beaus\'{e}jour}, \citenamefont
  {Delage}, \citenamefont {Leboeuf}, \citenamefont {Chang}, \citenamefont
  {Ramshaw}, \citenamefont {Bonn}, \citenamefont {Hardy}, \citenamefont
  {Liang}, \citenamefont {Adachi}, \citenamefont {Hussey}, \citenamefont
  {Vignolle}, \citenamefont {Proust}, \citenamefont {Sutherland}, \citenamefont
  {Kr\"{a}mer}, \citenamefont {Park}, \citenamefont {Graf}, \citenamefont
  {Doiron-Leyraud},\ and\ \citenamefont {Taillefer}}]{Grissonnanche2014a}%
  \BibitemOpen
  \bibfield  {author} {\bibinfo {author} {\bibfnamefont {G.}~\bibnamefont
  {Grissonnanche}}, \bibinfo {author} {\bibfnamefont {O.}~\bibnamefont
  {Cyr-Choini\`{e}re}}, \bibinfo {author} {\bibfnamefont {F.}~\bibnamefont
  {Lalibert\'{e}}}, \bibinfo {author} {\bibfnamefont {S.}~\bibnamefont
  {{Ren\'{e} de Cotret}}}, \bibinfo {author} {\bibfnamefont {A.}~\bibnamefont
  {Juneau-Fecteau}}, \bibinfo {author} {\bibfnamefont {S.}~\bibnamefont
  {Dufour-Beaus\'{e}jour}}, \bibinfo {author} {\bibfnamefont {M.-E.}\
  \bibnamefont {Delage}}, \bibinfo {author} {\bibfnamefont {D.}~\bibnamefont
  {Leboeuf}}, \bibinfo {author} {\bibfnamefont {J.}~\bibnamefont {Chang}},
  \bibinfo {author} {\bibfnamefont {B.~J.}\ \bibnamefont {Ramshaw}}, \bibinfo
  {author} {\bibfnamefont {D.~A.}\ \bibnamefont {Bonn}}, \bibinfo {author}
  {\bibfnamefont {W.~N.}\ \bibnamefont {Hardy}}, \bibinfo {author}
  {\bibfnamefont {R.}~\bibnamefont {Liang}}, \bibinfo {author} {\bibfnamefont
  {S.}~\bibnamefont {Adachi}}, \bibinfo {author} {\bibfnamefont {N.~E.}\
  \bibnamefont {Hussey}}, \bibinfo {author} {\bibfnamefont {B.}~\bibnamefont
  {Vignolle}}, \bibinfo {author} {\bibfnamefont {C.}~\bibnamefont {Proust}},
  \bibinfo {author} {\bibfnamefont {M.}~\bibnamefont {Sutherland}}, \bibinfo
  {author} {\bibfnamefont {S.}~\bibnamefont {Kr\"{a}mer}}, \bibinfo {author}
  {\bibfnamefont {J.-H.}\ \bibnamefont {Park}}, \bibinfo {author}
  {\bibfnamefont {D.}~\bibnamefont {Graf}}, \bibinfo {author} {\bibfnamefont
  {N.}~\bibnamefont {Doiron-Leyraud}}, \ and\ \bibinfo {author} {\bibfnamefont
  {L.}~\bibnamefont {Taillefer}},\ }\href {\doibase 10.1038/ncomms4280}
  {\bibfield  {journal} {\bibinfo  {journal} {Nature Communications}\ }\textbf
  {\bibinfo {volume} {5}},\ \bibinfo {pages} {3280} (\bibinfo {year}
  {2014})}\BibitemShut {NoStop}%
\bibitem [{\citenamefont {Riggs}\ \emph {et~al.}(2011)\citenamefont {Riggs},
  \citenamefont {Vafek}, \citenamefont {Kemper}, \citenamefont {Betts},
  \citenamefont {Migliori}, \citenamefont {Balakirev}, \citenamefont {Hardy},
  \citenamefont {Liang}, \citenamefont {Bonn},\ and\ \citenamefont
  {Boebinger}}]{Riggs2011}%
  \BibitemOpen
  \bibfield  {author} {\bibinfo {author} {\bibfnamefont {S.~C.}\ \bibnamefont
  {Riggs}}, \bibinfo {author} {\bibfnamefont {O.}~\bibnamefont {Vafek}},
  \bibinfo {author} {\bibfnamefont {J.~B.}\ \bibnamefont {Kemper}}, \bibinfo
  {author} {\bibfnamefont {J.~B.}\ \bibnamefont {Betts}}, \bibinfo {author}
  {\bibfnamefont {A.}~\bibnamefont {Migliori}}, \bibinfo {author}
  {\bibfnamefont {F.~F.}\ \bibnamefont {Balakirev}}, \bibinfo {author}
  {\bibfnamefont {W.~N.}\ \bibnamefont {Hardy}}, \bibinfo {author}
  {\bibfnamefont {R.}~\bibnamefont {Liang}}, \bibinfo {author} {\bibfnamefont
  {D.~A.}\ \bibnamefont {Bonn}}, \ and\ \bibinfo {author} {\bibfnamefont
  {G.~S.}\ \bibnamefont {Boebinger}},\ }\href {\doibase 10.1038/nphys1921}
  {\bibfield  {journal} {\bibinfo  {journal} {Nature Physics}\ }\textbf
  {\bibinfo {volume} {7}},\ \bibinfo {pages} {332} (\bibinfo {year}
  {2011})}\BibitemShut {NoStop}%
\bibitem [{\citenamefont {Ramshaw}\ \emph {et~al.}(2012)\citenamefont
  {Ramshaw}, \citenamefont {Day}, \citenamefont {Vignolle}, \citenamefont
  {LeBoeuf}, \citenamefont {Dosanjh}, \citenamefont {Proust}, \citenamefont
  {Taillefer}, \citenamefont {Liang}, \citenamefont {Hardy},\ and\
  \citenamefont {Bonn}}]{Ramshaw2012a}%
  \BibitemOpen
  \bibfield  {author} {\bibinfo {author} {\bibfnamefont {B.~J.}\ \bibnamefont
  {Ramshaw}}, \bibinfo {author} {\bibfnamefont {J.}~\bibnamefont {Day}},
  \bibinfo {author} {\bibfnamefont {B.}~\bibnamefont {Vignolle}}, \bibinfo
  {author} {\bibfnamefont {D.}~\bibnamefont {LeBoeuf}}, \bibinfo {author}
  {\bibfnamefont {P.}~\bibnamefont {Dosanjh}}, \bibinfo {author} {\bibfnamefont
  {C.}~\bibnamefont {Proust}}, \bibinfo {author} {\bibfnamefont
  {L.}~\bibnamefont {Taillefer}}, \bibinfo {author} {\bibfnamefont
  {R.}~\bibnamefont {Liang}}, \bibinfo {author} {\bibfnamefont {W.~N.}\
  \bibnamefont {Hardy}}, \ and\ \bibinfo {author} {\bibfnamefont {D.~A.}\
  \bibnamefont {Bonn}},\ }\href {\doibase 10.1103/PhysRevB.86.174501}
  {\bibfield  {journal} {\bibinfo  {journal} {Physical Review B}\ }\textbf
  {\bibinfo {volume} {86}},\ \bibinfo {pages} {174501} (\bibinfo {year}
  {2012})}\BibitemShut {NoStop}%
\bibitem [{\citenamefont {Chang}\ \emph {et~al.}(2012)\citenamefont {Chang},
  \citenamefont {Blackburn}, \citenamefont {Holmes}, \citenamefont
  {Christensen}, \citenamefont {Larsen}, \citenamefont {Mesot}, \citenamefont
  {Liang}, \citenamefont {Bonn}, \citenamefont {Hardy}, \citenamefont
  {Watenphul}, \citenamefont {Zimmermann}, \citenamefont {Forgan},\ and\
  \citenamefont {Hayden}}]{Chang2012}%
  \BibitemOpen
  \bibfield  {author} {\bibinfo {author} {\bibfnamefont {J.}~\bibnamefont
  {Chang}}, \bibinfo {author} {\bibfnamefont {E.}~\bibnamefont {Blackburn}},
  \bibinfo {author} {\bibfnamefont {A.~T.}\ \bibnamefont {Holmes}}, \bibinfo
  {author} {\bibfnamefont {N.~B.}\ \bibnamefont {Christensen}}, \bibinfo
  {author} {\bibfnamefont {J.}~\bibnamefont {Larsen}}, \bibinfo {author}
  {\bibfnamefont {J.}~\bibnamefont {Mesot}}, \bibinfo {author} {\bibfnamefont
  {R.}~\bibnamefont {Liang}}, \bibinfo {author} {\bibfnamefont {D.~A.}\
  \bibnamefont {Bonn}}, \bibinfo {author} {\bibfnamefont {W.~N.}\ \bibnamefont
  {Hardy}}, \bibinfo {author} {\bibfnamefont {A.}~\bibnamefont {Watenphul}},
  \bibinfo {author} {\bibfnamefont {M.~V.}\ \bibnamefont {Zimmermann}},
  \bibinfo {author} {\bibfnamefont {E.~M.}\ \bibnamefont {Forgan}}, \ and\
  \bibinfo {author} {\bibfnamefont {S.~M.}\ \bibnamefont {Hayden}},\ }\href
  {\doibase 10.1038/nphys2456} {\bibfield  {journal} {\bibinfo  {journal}
  {Nature Physics}\ }\textbf {\bibinfo {volume} {8}},\ \bibinfo {pages} {871}
  (\bibinfo {year} {2012})}\BibitemShut {NoStop}%
\bibitem [{\citenamefont {Gerber}\ \emph {et~al.}()\citenamefont {Gerber},
  \citenamefont {Jang}, \citenamefont {Nojiri}, \citenamefont {Matsuzawa},
  \citenamefont {Yasumura}, \citenamefont {Bonn}, \citenamefont {Liang},
  \citenamefont {Hardy}, \citenamefont {Islam}, \citenamefont {Mehta},
  \citenamefont {Song}, \citenamefont {Sikorski}, \citenamefont {Stefanescu},
  \citenamefont {Feng}, \citenamefont {Kivelson}, \citenamefont {Devereaux},
  \citenamefont {Shen}, \citenamefont {Kao}, \citenamefont {Lee}, \citenamefont
  {Zhu},\ and\ \citenamefont {Lee}}]{Gerber2015}%
  \BibitemOpen
  \bibfield  {author} {\bibinfo {author} {\bibfnamefont {S.}~\bibnamefont
  {Gerber}}, \bibinfo {author} {\bibfnamefont {H.}~\bibnamefont {Jang}},
  \bibinfo {author} {\bibfnamefont {H.}~\bibnamefont {Nojiri}}, \bibinfo
  {author} {\bibfnamefont {S.}~\bibnamefont {Matsuzawa}}, \bibinfo {author}
  {\bibfnamefont {H.}~\bibnamefont {Yasumura}}, \bibinfo {author}
  {\bibfnamefont {D.~A.}\ \bibnamefont {Bonn}}, \bibinfo {author}
  {\bibfnamefont {R.}~\bibnamefont {Liang}}, \bibinfo {author} {\bibfnamefont
  {W.~N.}\ \bibnamefont {Hardy}}, \bibinfo {author} {\bibfnamefont
  {Z.}~\bibnamefont {Islam}}, \bibinfo {author} {\bibfnamefont
  {A.}~\bibnamefont {Mehta}}, \bibinfo {author} {\bibfnamefont
  {S.}~\bibnamefont {Song}}, \bibinfo {author} {\bibfnamefont {M.}~\bibnamefont
  {Sikorski}}, \bibinfo {author} {\bibfnamefont {D.}~\bibnamefont
  {Stefanescu}}, \bibinfo {author} {\bibfnamefont {Y.}~\bibnamefont {Feng}},
  \bibinfo {author} {\bibfnamefont {S.~A.}\ \bibnamefont {Kivelson}}, \bibinfo
  {author} {\bibfnamefont {T.~P.}\ \bibnamefont {Devereaux}}, \bibinfo {author}
  {\bibfnamefont {Z.~X.}\ \bibnamefont {Shen}}, \bibinfo {author}
  {\bibfnamefont {C.~C.}\ \bibnamefont {Kao}}, \bibinfo {author} {\bibfnamefont
  {W.~S.}\ \bibnamefont {Lee}}, \bibinfo {author} {\bibfnamefont
  {D.}~\bibnamefont {Zhu}}, \ and\ \bibinfo {author} {\bibfnamefont {J.~S.}\
  \bibnamefont {Lee}},\ }\href {http://arxiv.org/abs/1506.07910} {\ }\Eprint
  {http://arxiv.org/abs/1506.07910} {arXiv:1506.07910} \BibitemShut {NoStop}%
\bibitem [{\citenamefont {Wu}\ \emph {et~al.}(2013)\citenamefont {Wu},
  \citenamefont {Mayaffre}, \citenamefont {Kr\"{a}mer}, \citenamefont
  {Horvati\'{c}}, \citenamefont {Berthier}, \citenamefont {Kuhns},
  \citenamefont {Reyes}, \citenamefont {Liang}, \citenamefont {Hardy},
  \citenamefont {Bonn},\ and\ \citenamefont {Julien}}]{Wu2013}%
  \BibitemOpen
  \bibfield  {author} {\bibinfo {author} {\bibfnamefont {T.}~\bibnamefont
  {Wu}}, \bibinfo {author} {\bibfnamefont {H.}~\bibnamefont {Mayaffre}},
  \bibinfo {author} {\bibfnamefont {S.}~\bibnamefont {Kr\"{a}mer}}, \bibinfo
  {author} {\bibfnamefont {M.}~\bibnamefont {Horvati\'{c}}}, \bibinfo {author}
  {\bibfnamefont {C.}~\bibnamefont {Berthier}}, \bibinfo {author}
  {\bibfnamefont {P.~L.}\ \bibnamefont {Kuhns}}, \bibinfo {author}
  {\bibfnamefont {A.~P.}\ \bibnamefont {Reyes}}, \bibinfo {author}
  {\bibfnamefont {R.}~\bibnamefont {Liang}}, \bibinfo {author} {\bibfnamefont
  {W.~N.}\ \bibnamefont {Hardy}}, \bibinfo {author} {\bibfnamefont {D.~A.}\
  \bibnamefont {Bonn}}, \ and\ \bibinfo {author} {\bibfnamefont {M.-H.}\
  \bibnamefont {Julien}},\ }\href {\doibase 10.1038/ncomms3113} {\bibfield
  {journal} {\bibinfo  {journal} {Nature Communications}\ }\textbf {\bibinfo
  {volume} {4}},\ \bibinfo {pages} {2113} (\bibinfo {year} {2013})}\BibitemShut
  {NoStop}%
\bibitem [{\citenamefont {LeBoeuf}\ \emph {et~al.}(2013)\citenamefont
  {LeBoeuf}, \citenamefont {Kr\"{a}mer}, \citenamefont {Hardy}, \citenamefont
  {Liang}, \citenamefont {Bonn},\ and\ \citenamefont {Proust}}]{LeBoeuf2012}%
  \BibitemOpen
  \bibfield  {author} {\bibinfo {author} {\bibfnamefont {D.}~\bibnamefont
  {LeBoeuf}}, \bibinfo {author} {\bibfnamefont {S.}~\bibnamefont {Kr\"{a}mer}},
  \bibinfo {author} {\bibfnamefont {W.~N.}\ \bibnamefont {Hardy}}, \bibinfo
  {author} {\bibfnamefont {R.}~\bibnamefont {Liang}}, \bibinfo {author}
  {\bibfnamefont {D.~A.}\ \bibnamefont {Bonn}}, \ and\ \bibinfo {author}
  {\bibfnamefont {C.}~\bibnamefont {Proust}},\ }\href {\doibase
  10.1038/nphys2502} {\bibfield  {journal} {\bibinfo  {journal} {Nature
  Physics}\ }\textbf {\bibinfo {volume} {9}},\ \bibinfo {pages} {79} (\bibinfo
  {year} {2013})}\BibitemShut {NoStop}%
\bibitem [{\citenamefont {H\"{u}cker}\ \emph {et~al.}(2014)\citenamefont
  {H\"{u}cker}, \citenamefont {Christensen}, \citenamefont {Holmes},
  \citenamefont {Blackburn}, \citenamefont {Forgan}, \citenamefont {Liang},
  \citenamefont {Bonn}, \citenamefont {Hardy}, \citenamefont {Gutowski},
  \citenamefont {Zimmermann}, \citenamefont {Hayden},\ and\ \citenamefont
  {Chang}}]{Hucker2014}%
  \BibitemOpen
  \bibfield  {author} {\bibinfo {author} {\bibfnamefont {M.}~\bibnamefont
  {H\"{u}cker}}, \bibinfo {author} {\bibfnamefont {N.~B.}\ \bibnamefont
  {Christensen}}, \bibinfo {author} {\bibfnamefont {A.~T.}\ \bibnamefont
  {Holmes}}, \bibinfo {author} {\bibfnamefont {E.}~\bibnamefont {Blackburn}},
  \bibinfo {author} {\bibfnamefont {E.~M.}\ \bibnamefont {Forgan}}, \bibinfo
  {author} {\bibfnamefont {R.}~\bibnamefont {Liang}}, \bibinfo {author}
  {\bibfnamefont {D.~A.}\ \bibnamefont {Bonn}}, \bibinfo {author}
  {\bibfnamefont {W.~N.}\ \bibnamefont {Hardy}}, \bibinfo {author}
  {\bibfnamefont {O.}~\bibnamefont {Gutowski}}, \bibinfo {author}
  {\bibfnamefont {M.~V.}\ \bibnamefont {Zimmermann}}, \bibinfo {author}
  {\bibfnamefont {S.~M.}\ \bibnamefont {Hayden}}, \ and\ \bibinfo {author}
  {\bibfnamefont {J.}~\bibnamefont {Chang}},\ }\href@noop {} {\bibfield
  {journal} {\bibinfo  {journal} {Physical Review B}\ }\textbf {\bibinfo
  {volume} {90}},\ \bibinfo {pages} {054514} (\bibinfo {year}
  {2014})}\BibitemShut {NoStop}%
\bibitem [{\citenamefont {Marcenat}\ \emph {et~al.}(2015)\citenamefont
  {Marcenat}, \citenamefont {Demuer}, \citenamefont {Beauvois}, \citenamefont
  {Michon}, \citenamefont {Grockowiak}, \citenamefont {Liang}, \citenamefont
  {Hardy}, \citenamefont {Bonn},\ and\ \citenamefont {Klein}}]{Marcenat2015}%
  \BibitemOpen
  \bibfield  {author} {\bibinfo {author} {\bibfnamefont {C.}~\bibnamefont
  {Marcenat}}, \bibinfo {author} {\bibfnamefont {A.}~\bibnamefont {Demuer}},
  \bibinfo {author} {\bibfnamefont {K.}~\bibnamefont {Beauvois}}, \bibinfo
  {author} {\bibfnamefont {B.}~\bibnamefont {Michon}}, \bibinfo {author}
  {\bibfnamefont {A.}~\bibnamefont {Grockowiak}}, \bibinfo {author}
  {\bibfnamefont {R.}~\bibnamefont {Liang}}, \bibinfo {author} {\bibfnamefont
  {W.}~\bibnamefont {Hardy}}, \bibinfo {author} {\bibfnamefont {D.~A.}\
  \bibnamefont {Bonn}}, \ and\ \bibinfo {author} {\bibfnamefont
  {T.}~\bibnamefont {Klein}},\ }\href {\doibase 10.1038/ncomms8927} {\bibfield
  {journal} {\bibinfo  {journal} {Nature Communications}\ }\textbf {\bibinfo
  {volume} {6}},\ \bibinfo {pages} {7927} (\bibinfo {year} {2015})}\BibitemShut
  {NoStop}%
\bibitem [{\citenamefont {Kokanovi\'{c}}\ \emph {et~al.}(2012)\citenamefont
  {Kokanovi\'{c}}, \citenamefont {Cooper},\ and\ \citenamefont
  {Iida}}]{Kokanovic2012}%
  \BibitemOpen
  \bibfield  {author} {\bibinfo {author} {\bibfnamefont {I.}~\bibnamefont
  {Kokanovi\'{c}}}, \bibinfo {author} {\bibfnamefont {J.~R.}\ \bibnamefont
  {Cooper}}, \ and\ \bibinfo {author} {\bibfnamefont {K.}~\bibnamefont
  {Iida}},\ }\href {\doibase 10.1209/0295-5075/98/57011} {\bibfield  {journal}
  {\bibinfo  {journal} {Europhysics Letters}\ }\textbf {\bibinfo {volume}
  {98}},\ \bibinfo {pages} {57011} (\bibinfo {year} {2012})}\BibitemShut
  {NoStop}%
\bibitem [{\citenamefont {Kokanovi\'{c}}\ \emph {et~al.}(2013)\citenamefont
  {Kokanovi\'{c}}, \citenamefont {Hills}, \citenamefont {Sutherland},
  \citenamefont {Liang},\ and\ \citenamefont {Cooper}}]{Kokanovic2013a}%
  \BibitemOpen
  \bibfield  {author} {\bibinfo {author} {\bibfnamefont {I.}~\bibnamefont
  {Kokanovi\'{c}}}, \bibinfo {author} {\bibfnamefont {D.~J.}\ \bibnamefont
  {Hills}}, \bibinfo {author} {\bibfnamefont {M.~L.}\ \bibnamefont
  {Sutherland}}, \bibinfo {author} {\bibfnamefont {R.}~\bibnamefont {Liang}}, \
  and\ \bibinfo {author} {\bibfnamefont {J.~R.}\ \bibnamefont {Cooper}},\
  }\href {\doibase 10.1103/PhysRevB.88.060505} {\bibfield  {journal} {\bibinfo
  {journal} {Physical Review B}\ }\textbf {\bibinfo {volume} {88}},\ \bibinfo
  {pages} {060505} (\bibinfo {year} {2013})}\BibitemShut {NoStop}%
\bibitem [{\citenamefont {Loram}\ \emph {et~al.}(1998)\citenamefont {Loram},
  \citenamefont {Mirza}, \citenamefont {Cooper},\ and\ \citenamefont
  {Tallon}}]{Loram1998}%
  \BibitemOpen
  \bibfield  {author} {\bibinfo {author} {\bibfnamefont {J.~W.}\ \bibnamefont
  {Loram}}, \bibinfo {author} {\bibfnamefont {K.~A.}\ \bibnamefont {Mirza}},
  \bibinfo {author} {\bibfnamefont {J.~R.}\ \bibnamefont {Cooper}}, \ and\
  \bibinfo {author} {\bibfnamefont {J.~L.}\ \bibnamefont {Tallon}},\ }\href
  {\doibase 10.1016/S0022-3697(98)00180-2} {\bibfield  {journal} {\bibinfo
  {journal} {Journal of Physics and Chemistry of Solids}\ }\textbf {\bibinfo
  {volume} {59}},\ \bibinfo {pages} {2091} (\bibinfo {year}
  {1998})}\BibitemShut {NoStop}%
\bibitem [{Note2()}]{Note2}%
  \BibitemOpen
  \bibinfo {note} {For OII crystal, the calibration factor used to obtain the
  absolute value of $M_c$ was found by requiring the $B^2$ term in the torque
  data ($\tau /V = 1/2\chi B^2\protect \qopname \relax o{sin}2\theta $) between
  20 and 64 T at 80 K to give the value $\chi _D(80) = 11.2$ A/m/T on the blue
  line in Fig. \ref {fig:BG}. For OVIII, excellent $\protect \qopname \relax
  o{sin}2\theta $ behaviour was observed in the torque at 10 T and 97 K at the
  HFML, and was then normalized to the measured value of $\chi _D(97)$ obtained
  from angle sweeps on the same crystal in the Cavendish Laboratory at
  Cambridge, whose results are shown by solid points in Fig. \ref {fig:BG}. In
  the present high field experiments the crystals had to be small, so the
  $T$-dependence of the lever sensitivity could not be found from the
  gravitational torque in the usual way. It was obtained from the consistent
  normalized $T$-dependences of several levers of the same type in zero field
  gravitational torque measurements on larger crystals.}\BibitemShut {Stop}%
\bibitem [{\citenamefont {Hao}\ and\ \citenamefont
  {Clem}(1991{\natexlab{a}})}]{Hao1991a}%
  \BibitemOpen
  \bibfield  {author} {\bibinfo {author} {\bibfnamefont {Z.}~\bibnamefont
  {Hao}}\ and\ \bibinfo {author} {\bibfnamefont {J.~R.}\ \bibnamefont {Clem}},\
  }\href {\doibase 10.1103/PhysRevLett.67.2371} {\bibfield  {journal} {\bibinfo
   {journal} {Physical Review Letters}\ }\textbf {\bibinfo {volume} {67}},\
  \bibinfo {pages} {2371} (\bibinfo {year} {1991}{\natexlab{a}})}\BibitemShut
  {NoStop}%
\bibitem [{\citenamefont {Sonier}\ \emph {et~al.}(2007)\citenamefont {Sonier},
  \citenamefont {Sabok-Sayr}, \citenamefont {Callaghan}, \citenamefont
  {Kaiser}, \citenamefont {Pacradouni}, \citenamefont {Brewer}, \citenamefont
  {Stubbs}, \citenamefont {Hardy}, \citenamefont {Bonn}, \citenamefont
  {Liang},\ and\ \citenamefont {Atkinson}}]{Sonier2007}%
  \BibitemOpen
  \bibfield  {author} {\bibinfo {author} {\bibfnamefont {J.~E.}\ \bibnamefont
  {Sonier}}, \bibinfo {author} {\bibfnamefont {S.~A.}\ \bibnamefont
  {Sabok-Sayr}}, \bibinfo {author} {\bibfnamefont {F.~D.}\ \bibnamefont
  {Callaghan}}, \bibinfo {author} {\bibfnamefont {C.~V.}\ \bibnamefont
  {Kaiser}}, \bibinfo {author} {\bibfnamefont {V.}~\bibnamefont {Pacradouni}},
  \bibinfo {author} {\bibfnamefont {J.~H.}\ \bibnamefont {Brewer}}, \bibinfo
  {author} {\bibfnamefont {S.~L.}\ \bibnamefont {Stubbs}}, \bibinfo {author}
  {\bibfnamefont {W.~N.}\ \bibnamefont {Hardy}}, \bibinfo {author}
  {\bibfnamefont {D.~A.}\ \bibnamefont {Bonn}}, \bibinfo {author}
  {\bibfnamefont {R.}~\bibnamefont {Liang}}, \ and\ \bibinfo {author}
  {\bibfnamefont {W.~A.}\ \bibnamefont {Atkinson}},\ }\href {\doibase
  10.1103/PhysRevB.76.134518} {\bibfield  {journal} {\bibinfo  {journal}
  {Physical Review B}\ }\textbf {\bibinfo {volume} {76}},\ \bibinfo {pages}
  {134518} (\bibinfo {year} {2007})}\BibitemShut {NoStop}%
\bibitem [{\citenamefont {Benseman}\ \emph {et~al.}()\citenamefont {Benseman},
  \citenamefont {Cooper},\ and\ \citenamefont {Balakrishnan}}]{Benseman2015}%
  \BibitemOpen
  \bibfield  {author} {\bibinfo {author} {\bibfnamefont {T.~M.}\ \bibnamefont
  {Benseman}}, \bibinfo {author} {\bibfnamefont {J.~R.}\ \bibnamefont
  {Cooper}}, \ and\ \bibinfo {author} {\bibfnamefont {G.}~\bibnamefont
  {Balakrishnan}},\ }\href {http://arxiv.org/abs/1503.00335} {\ }\Eprint
  {http://arxiv.org/abs/1503.00335} {arXiv:1503.00335} \BibitemShut {NoStop}%
\bibitem [{\citenamefont {Pogosov}\ \emph {et~al.}(2001)\citenamefont
  {Pogosov}, \citenamefont {Kugel}, \citenamefont {Rakhmanov},\ and\
  \citenamefont {Brandt}}]{Pogosov2000}%
  \BibitemOpen
  \bibfield  {author} {\bibinfo {author} {\bibfnamefont {W.~V.}\ \bibnamefont
  {Pogosov}}, \bibinfo {author} {\bibfnamefont {K.~I.}\ \bibnamefont {Kugel}},
  \bibinfo {author} {\bibfnamefont {A.~L.}\ \bibnamefont {Rakhmanov}}, \ and\
  \bibinfo {author} {\bibfnamefont {E.~H.}\ \bibnamefont {Brandt}},\ }\href
  {\doibase 10.1103/PhysRevB.64.064517} {\bibfield  {journal} {\bibinfo
  {journal} {Physical Review B}\ }\textbf {\bibinfo {volume} {64}},\ \bibinfo
  {pages} {064517} (\bibinfo {year} {2001})}\BibitemShut {NoStop}%
\bibitem [{\citenamefont {Bosma}\ \emph {et~al.}(2011)\citenamefont {Bosma},
  \citenamefont {Weyeneth}, \citenamefont {Puzniak}, \citenamefont {Erb},
  \citenamefont {Schilling},\ and\ \citenamefont {Keller}}]{Bosma2011}%
  \BibitemOpen
  \bibfield  {author} {\bibinfo {author} {\bibfnamefont {S.}~\bibnamefont
  {Bosma}}, \bibinfo {author} {\bibfnamefont {S.}~\bibnamefont {Weyeneth}},
  \bibinfo {author} {\bibfnamefont {R.}~\bibnamefont {Puzniak}}, \bibinfo
  {author} {\bibfnamefont {A.}~\bibnamefont {Erb}}, \bibinfo {author}
  {\bibfnamefont {A.}~\bibnamefont {Schilling}}, \ and\ \bibinfo {author}
  {\bibfnamefont {H.}~\bibnamefont {Keller}},\ }\href {\doibase
  10.1103/PhysRevB.84.024514} {\bibfield  {journal} {\bibinfo  {journal}
  {Physical Review B}\ }\textbf {\bibinfo {volume} {84}},\ \bibinfo {pages}
  {024514} (\bibinfo {year} {2011})}\BibitemShut {NoStop}%
\bibitem [{\citenamefont {Tinkham}(1996)}]{Tinkham1996}%
  \BibitemOpen
  \bibfield  {author} {\bibinfo {author} {\bibfnamefont {M.}~\bibnamefont
  {Tinkham}},\ }\href@noop {} {\emph {\bibinfo {title} {Introduction to
  {S}uperconductivity}}},\ \bibinfo {edition} {2nd}\ ed.\ (\bibinfo
  {publisher} {Dover},\ \bibinfo {address} {Mineola, New York},\ \bibinfo
  {year} {1996})\BibitemShut {NoStop}%
\bibitem [{\citenamefont {Hao}\ and\ \citenamefont
  {Clem}(1991{\natexlab{b}})}]{Hao1991}%
  \BibitemOpen
  \bibfield  {author} {\bibinfo {author} {\bibfnamefont {Z.}~\bibnamefont
  {Hao}}\ and\ \bibinfo {author} {\bibfnamefont {J.~R.}\ \bibnamefont {Clem}},\
  }\href {\doibase 10.1103/PhysRevB.43.7622} {\bibfield  {journal} {\bibinfo
  {journal} {Physical Review B}\ }\textbf {\bibinfo {volume} {43}},\ \bibinfo
  {pages} {7622} (\bibinfo {year} {1991}{\natexlab{b}})}\BibitemShut {NoStop}%
\bibitem [{\citenamefont {LeBoeuf}\ \emph {et~al.}(2007)\citenamefont
  {LeBoeuf}, \citenamefont {Doiron-Leyraud}, \citenamefont {Levallois},
  \citenamefont {Daou}, \citenamefont {Bonnemaison}, \citenamefont {Hussey},
  \citenamefont {Balicas}, \citenamefont {Ramshaw}, \citenamefont {Liang},
  \citenamefont {Bonn}, \citenamefont {Hardy}, \citenamefont {Adachi},
  \citenamefont {Proust},\ and\ \citenamefont {Taillefer}}]{LeBoeuf2007}%
  \BibitemOpen
  \bibfield  {author} {\bibinfo {author} {\bibfnamefont {D.}~\bibnamefont
  {LeBoeuf}}, \bibinfo {author} {\bibfnamefont {N.}~\bibnamefont
  {Doiron-Leyraud}}, \bibinfo {author} {\bibfnamefont {J.}~\bibnamefont
  {Levallois}}, \bibinfo {author} {\bibfnamefont {R.}~\bibnamefont {Daou}},
  \bibinfo {author} {\bibfnamefont {J.-B.}\ \bibnamefont {Bonnemaison}},
  \bibinfo {author} {\bibfnamefont {N.~E.}\ \bibnamefont {Hussey}}, \bibinfo
  {author} {\bibfnamefont {L.}~\bibnamefont {Balicas}}, \bibinfo {author}
  {\bibfnamefont {B.~J.}\ \bibnamefont {Ramshaw}}, \bibinfo {author}
  {\bibfnamefont {R.}~\bibnamefont {Liang}}, \bibinfo {author} {\bibfnamefont
  {D.~A.}\ \bibnamefont {Bonn}}, \bibinfo {author} {\bibfnamefont {W.~N.}\
  \bibnamefont {Hardy}}, \bibinfo {author} {\bibfnamefont {S.}~\bibnamefont
  {Adachi}}, \bibinfo {author} {\bibfnamefont {C.}~\bibnamefont {Proust}}, \
  and\ \bibinfo {author} {\bibfnamefont {L.}~\bibnamefont {Taillefer}},\ }\href
  {\doibase 10.1038/nature06332} {\bibfield  {journal} {\bibinfo  {journal}
  {Nature}\ }\textbf {\bibinfo {volume} {450}},\ \bibinfo {pages} {533}
  (\bibinfo {year} {2007})}\BibitemShut {NoStop}%
\bibitem [{\citenamefont {Galitski}\ and\ \citenamefont
  {Larkin}(2001)}]{Galitski2000}%
  \BibitemOpen
  \bibfield  {author} {\bibinfo {author} {\bibfnamefont {V.~M.}\ \bibnamefont
  {Galitski}}\ and\ \bibinfo {author} {\bibfnamefont {A.~I.}\ \bibnamefont
  {Larkin}},\ }\href {\doibase 10.1103/PhysRevB.63.174506} {\bibfield
  {journal} {\bibinfo  {journal} {Physical Review B}\ }\textbf {\bibinfo
  {volume} {63}},\ \bibinfo {pages} {174506} (\bibinfo {year}
  {2001})}\BibitemShut {NoStop}%
\bibitem [{\citenamefont {Chang}\ \emph {et~al.}(2011)\citenamefont {Chang},
  \citenamefont {Doiron-Leyraud}, \citenamefont {Lalibert\'{e}}, \citenamefont
  {Daou}, \citenamefont {LeBoeuf}, \citenamefont {Ramshaw}, \citenamefont
  {Liang}, \citenamefont {Bonn}, \citenamefont {Hardy}, \citenamefont {Proust},
  \citenamefont {Sheikin}, \citenamefont {Behnia},\ and\ \citenamefont
  {Taillefer}}]{Chang2011}%
  \BibitemOpen
  \bibfield  {author} {\bibinfo {author} {\bibfnamefont {J.}~\bibnamefont
  {Chang}}, \bibinfo {author} {\bibfnamefont {N.}~\bibnamefont
  {Doiron-Leyraud}}, \bibinfo {author} {\bibfnamefont {F.}~\bibnamefont
  {Lalibert\'{e}}}, \bibinfo {author} {\bibfnamefont {R.}~\bibnamefont {Daou}},
  \bibinfo {author} {\bibfnamefont {D.}~\bibnamefont {LeBoeuf}}, \bibinfo
  {author} {\bibfnamefont {B.~J.}\ \bibnamefont {Ramshaw}}, \bibinfo {author}
  {\bibfnamefont {R.}~\bibnamefont {Liang}}, \bibinfo {author} {\bibfnamefont
  {D.~A.}\ \bibnamefont {Bonn}}, \bibinfo {author} {\bibfnamefont {W.~N.}\
  \bibnamefont {Hardy}}, \bibinfo {author} {\bibfnamefont {C.}~\bibnamefont
  {Proust}}, \bibinfo {author} {\bibfnamefont {I.}~\bibnamefont {Sheikin}},
  \bibinfo {author} {\bibfnamefont {K.}~\bibnamefont {Behnia}}, \ and\ \bibinfo
  {author} {\bibfnamefont {L.}~\bibnamefont {Taillefer}},\ }\href {\doibase
  10.1103/PhysRevB.84.014507} {\bibfield  {journal} {\bibinfo  {journal}
  {Physical Review B}\ }\textbf {\bibinfo {volume} {84}},\ \bibinfo {pages}
  {014507} (\bibinfo {year} {2011})}\BibitemShut {NoStop}%
\bibitem [{\citenamefont {Hayward}\ \emph {et~al.}(2014)\citenamefont
  {Hayward}, \citenamefont {Achkar}, \citenamefont {Hawthorn}, \citenamefont
  {Melko},\ and\ \citenamefont {Sachdev}}]{Hayward2014a}%
  \BibitemOpen
  \bibfield  {author} {\bibinfo {author} {\bibfnamefont {L.~E.}\ \bibnamefont
  {Hayward}}, \bibinfo {author} {\bibfnamefont {A.~J.}\ \bibnamefont {Achkar}},
  \bibinfo {author} {\bibfnamefont {D.~G.}\ \bibnamefont {Hawthorn}}, \bibinfo
  {author} {\bibfnamefont {R.~G.}\ \bibnamefont {Melko}}, \ and\ \bibinfo
  {author} {\bibfnamefont {S.}~\bibnamefont {Sachdev}},\ }\href {\doibase
  10.1103/PhysRevB.90.094515} {\bibfield  {journal} {\bibinfo  {journal}
  {Physical Review B}\ }\textbf {\bibinfo {volume} {90}},\ \bibinfo {pages}
  {094515} (\bibinfo {year} {2014})}\BibitemShut {NoStop}%
\end{thebibliography}

\end{document}